# The Klein-Gordon Equation of Generalized Hulthen Potential in Complex Quantum Mechanics

Mehmet Simsek [1,a] and Harun Egrifes [2,b]

[1] Department of Physics, Faculty of Arts and Sciences, Gazi University, 06500 Ankara, Turkey

[2] Department of Physics, Faculty of Science, Ege University, 35100 Izmir, Turkey

**Abstract**

We have investigated the reality of exact bound states of complex and/or *PT*-symmetric non-Hermitian exponential-type generalized Hulthen potential. The Klein-Gordon equation has been solved by using the Nikiforov-Uvarov method which is based on solving the second-order linear differential equations by reduction to a generalized equation of hypergeometric type. In many cases of interest, negative and positive energy states have been discussed for different types of complex potentials.

[a] E-mail : msimsek@quark.fef.gazi.edu.tr

[b] E-mail : egrifes@sci.ege.edu.tr

# 1. Introduction

More recently, to overcome the weaker conditions of *PT*-symmetry Bender, Brody and Jones [1] emphasized new conditions in the context of complex quantum mechanics. Furthermore, in their study, including the standard condition of Hermiticity, the converse of the fundamental *CPT*-symmetry theorem [2] have been established with a charge-conjugation operator *C*. According to the new theory [1], if *CPT*-symmetry is not spontaneously broken, the eigenvalues of the observable are real. Following the early studies of Bender and his co-workers [3], the *PT*-symmetric formulation has been succesfully utilized by many authors [4-12]. The *PT*-symmetric but non-Hermitian Hamiltonians have real spectra whether the Hamiltonians are Hermitian or not. Non-Hermitian Hamiltonians with real or complex spectra have also been analyzed by using different methods [5-8,12-14].

In the view of *PT*-symmetric formulation, we will apply the Nikiforov-Uvarov ( *NU*-) method [15] to solve the Klein-Gordon (KG) equation. We have presented exact bound states for a family of exponential potentials, i.e., generalized Hulthen potential which is reducible to the standard Hulthen potential, Woods-Saxon potential, exponential-type screened potential. Note that using the quantization of the boundary condition of the states at the origin, Znojil [16] studied another generalized Hulthen and other exponential potentials in non-relativistic and relativistic regions. Also, Dominguez-Adame [17] and Chetouani *et al*. [18] studied relativistic bound states of standard Hulthen potential. We have also pointed out why the *NU*-method could not be applicable to the exponential-type potential. On the other hand, Rao and Kagali [19] investigated the relativistic bound states of the exponential potential by means of one-dimensional KG equation, and Znojil [20] found the non-relativistic solutions of Schrödinger equation. However, it is well known that for the exponential potential there is no explicit form of the energy expression of bound states for Schrödinger [21], KG [19] and also Dirac [22] equations.

We show that, it is possible to get relativistic representation of the *PT*-symmetric quantum mechanical formulation.

## 2. The Klein-Gordon Equation of *PT*-Symmetric Potentials

The KG equation for the free particle is [23], in natural units $\hbar = c = 1$,

$$\left( \frac{\partial}{\partial x_\mu} \frac{\partial}{\partial x^\mu} + m^2 \right) \psi = 0. \tag{1}$$

Defining $\psi = \theta + \chi$, $i\partial\psi/\partial t = (\theta - \chi)m$ [23], we can write the one-dimensional Schrödinger equation-like KG equation in the two components

$$i\frac{\partial \Phi(x)}{\partial t} = \left\{ \begin{bmatrix} 1 & 1 \\ -1 & -1 \end{bmatrix} \frac{\wp^2}{2m} + \begin{bmatrix} 1 & 0 \\ 0 & -1 \end{bmatrix} m + eV(x) \right\} \Phi(x) \quad (2)$$

for a spin-zero particle, in the scalar potential field $V(x)$, where $\wp = p - eA$. Note that, since the kinetic energy term involves non-Hermitian matrix, $\begin{bmatrix} 1 & 1 \\ -1 & -1 \end{bmatrix}$, the Schrödinger-like Hamiltonian is non-Hermitian and also non-*PT*-symmetric for any given $V(x)$. It is contrary to the Schrödinger equation for *PT*-symmetric and Hermitian Hamiltonians. Although the Hamiltonian which is given by Eq.(2) is non-Hermitian, the transformed one is:

$$H' = e^{iS} H e^{-iS} = \eta \sqrt{p^2 + m^2}, \quad (3)$$

where $\eta$ is the diagonal matrix $\begin{bmatrix} 1 & 0 \\ 0 & -1 \end{bmatrix}$ and $S = -\frac{i}{2} \begin{bmatrix} 0 & 1 \\ 1 & 0 \end{bmatrix} \tanh^{-1}\left( \frac{p^2/2m}{m + p^2/2m} \right)$ with $p = -i\nabla$.

By using Foldy-Wouthuysen approach [23], for the special case of static external fields, one can find the approximate Schrödinger equation up to order $1/m^4$ as,

$$i\frac{\partial \Phi'}{\partial t} = H'\Phi', \quad \Phi' = e^{iS}\Phi \quad (4)$$

with

$$H' = \eta\left(m + \frac{\wp^2}{2m} - \frac{\wp^4}{8m^3} + ...\right) + eV(x) + \frac{1}{32m^4}\left[\wp^2, \left[\wp^2, eV(x)\right]\right] + ... \quad (5)$$

In the case of non-relativistic quantum mechanics we have specified minimal coupling of electromagnetic field, $V(x)$,

$$E \rightarrow i\hbar \frac{\partial}{\partial t} - eV(x). \quad (6)$$

On the other hand, we will be dealing with bound state solutions, i.e., the wave function vanishes at infinity. For any given one-dimensional potential, one can arrive to the time-independent KG equation in the form of Schrödinger equation as [24]

$$\left\{ \frac{d^2}{dx^2} + \frac{1}{\hbar^2 c^2}\left[ (E - V(x))^2 - (mc^2)^2 \right] \right\} \psi(x) = 0 \quad (7)$$

where *V(x)* is the one-dimensional vector potential.

Let the potential be written as a complex potential

$$V(x) = V_R(x) + iV_I(x). \tag{8}$$

The potential is called *PT*-symmetric when

$$PTV(x) = V(x)PT, \tag{9}$$

i.e., *PT*-symmetry condition for a given potential *V(x)* reads

$$V^*(-x) = V(x). \tag{10}$$

## 3. The Nikiforov-Uvarov method

The non-relativistic Schrödinger equation and other Schrödinger-like equations can be solved by using *NU*-method which is based on the solutions of general second-order linear differential equation with special orthogonal functions [15]. It is well known that for any given one-dimensional or radial potential, the Schrödinger equation can be written as a second-order linear differential equation. However, in the *NU*-method generalized second-order linear differential equation can be written as,

$$\psi''(z) + \frac{\tilde{\tau}(z)}{\sigma(z)}\psi'(z) + \frac{\tilde{\sigma}(z)}{\sigma^2(z)}\psi(z) = 0 \tag{11}$$

where $\sigma(z)$ and $\tilde{\sigma}(z)$ are polynomials of degree at most two, and $\tilde{\tau}(z)$ is a polynomial of degree at most one.

Using the transformation

$$\psi(z) = \phi(z)y(z) \tag{12}$$

Eq.(11) could be reduced to the hypergeometric-type equation

$$\sigma(z)y'' + \tau(z)y' + \lambda y = 0 \tag{13}$$

whose polynomial solutions are given by the Rodrigues relation

$$y(z, \lambda_n) = y_n(z) \approx \frac{1}{\rho(z)} \frac{d^n}{dz^n}[\sigma^n(z)\rho(z)] \quad (n = 0,1,2,...) \tag{14}$$

where the weight function $\rho(z)$ satisfies the equation

$$\frac{d}{dz}[\sigma(z)\rho(z)] = \tau(z)\rho(z). \tag{15}$$

On the other hand, the function $\phi(z)$ satisfies the relation

$$\phi'(z)/\phi(z) = \pi(z)/\sigma(z) \tag{16}$$

with an arbitrary linear polynomial $\pi(z)$.

## 4. The Bound States of Generalized Hulthen Potential

We shall study the time-independent KG equation with a family of exponential potentials,

$$V_q(x) = -V_0 \frac{e^{-\alpha x}}{1 - q e^{-\alpha x}} \qquad (17)$$

which is called generalized Hulthen potential [25]. We have to note that, for some specific $q$ this potential reduces to the well-known types : such as for $q = 0$, to the exponential potential; for $q = 1$ to the standard Hulthen potential; and for $q = -1$ to the Woods-Saxon potential, respectively. The generalized Hulthen potential for different generalization parameters is illustrated in Fig.1.

In order to apply the *NU*-method, we can write the one-dimensional KG equation as a second-order linear differential equation for the generalized Hulthen potential,

$$\psi_q''(x) + \frac{1}{\hbar^2 c^2}\left[E^2 + 2V_0 E \frac{e^{-\alpha x}}{1 - qe^{-\alpha x}} + V_0^2 \frac{e^{-2\alpha x}}{(1 - qe^{-\alpha x})^2} - (mc^2)^2\right]\psi_q(x) = 0 \qquad (18)$$

by defining within a new variable $z = e^{-\alpha x}$, this equation is reduced to the generalized equation of hypergeometric type which is given by Eq.(11)

$$\psi_q''(z) + \frac{1 - qz}{z(1 - qz)}\psi_q'(z) + \frac{1}{[z(1 - qz)]^2}\left\{(\gamma^2 - q^2 \mathcal{E}^2 - q\beta^2)z^2 + (\beta^2 + 2q\mathcal{E}^2)z - \mathcal{E}^2\right\}\psi_q(z) = 0 \qquad (19)$$

with $\tilde{\tau}(z) = 1 - qz$, $\sigma(z) = z(1 - qz)$, $\tilde{\sigma}(z) = (\gamma^2 - q^2 \mathcal{E}^2 - q\beta^2)z^2 + (\beta^2 + 2q\mathcal{E}^2)z - \mathcal{E}^2$. We use some dimensionless abbreviations as

$$\mathcal{E}^2 = -\frac{1}{\hbar^2 c^2 \alpha^2}(E^2 - m^2 c^4), \quad \beta^2 = \frac{2V_0 E}{\hbar^2 c^2 \alpha^2}, \quad \gamma^2 = \frac{V_0^2}{\hbar^2 c^2 \alpha^2} \qquad (20)$$

with real $\mathcal{E} \geq 0$ ($E^2 \leq m^2 c^4$) for bound states.

In the *NU*-method the new function $\pi(z)$ is defined as

$$\pi(z) = \frac{\sigma'(z) - \tilde{\tau}(z)}{2} \pm \sqrt{\left(\frac{\sigma'(z) - \tilde{\tau}(z)}{2}\right)^2 - \tilde{\sigma}(z) + k\sigma(z)} \qquad (21)$$

and in the present case the function appears as

$$\pi(z) = -\frac{qz}{2} \pm \frac{1}{2}\sqrt{[q^2 - 4(\gamma^2 - q^2\mathcal{E}^2 - q\beta^2) - 4qk]z^2 + 4[k - (\beta^2 + 2q\mathcal{E}^2)]z + 4\mathcal{E}^2}. \qquad (22)$$

The constant parameter $k$ can be found by the condition that the expression under the square root has a double zero, i.e., its discriminant is zero. So, there are two possible functions for each $k$ :

$$\pi(z) = -\frac{qz}{2} \pm \begin{cases} \frac{1}{2}\left[(2q\mathcal{E} - a)z - 2\mathcal{E}\right] & \text{for } k = \beta^2 + \mathcal{E}a \\ \frac{1}{2}\left[(2q\mathcal{E} + a)z - 2\mathcal{E}\right] & \text{for } k = \beta^2 - \mathcal{E}a \end{cases} \quad (23)$$

where $a = \sqrt{q^2 - 4\gamma^2}$. According to the *NU*-method, with an appropriate choice of the function $\pi(z)$, $\pi(z) = \mathcal{E} - \frac{1}{2}\left[q + (2q\mathcal{E} + a)\right]z$ for $k = \beta^2 - \mathcal{E}a$, we can define a new function $\tau(z) = \tilde{\tau}(z) + 2\pi(z)$ which has a negative derivative and given by

$$\tau(z) = (1 + 2\mathcal{E}) - \left[2q + (2q\mathcal{E} + a)\right]z. \quad (24)$$

Then, we have another constant, $\lambda = k + \pi'(z)$, written as

$$\lambda = \beta^2 - \mathcal{E}a - \frac{1}{2}\left[q + (2q\mathcal{E} + a)\right]. \quad (25)$$

A new eigenvalue equation for a given $\lambda$, $\tau(z)$ and $\sigma(z)$ are defined as [15]

$$\lambda = \lambda_n = -n\tau' - \frac{n(n-1)}{2}\sigma''. \quad (n = 0,1,2,...) \quad (26)$$

Thus, substituting $\lambda$, $\tau'$ and $\sigma''$ in Eq.(26), the exact energy eigenvalues of the generalized Hulthen potential are determined as

$$E_n(V_0, q, \alpha) = \frac{V_0}{2q} \pm \kappa \sqrt{\frac{(mc^2)^2}{4\gamma^2 + \kappa^2} - \frac{V_0^2}{16q^2\gamma^2}} \quad (27)$$

where $\kappa = \sqrt{q^2 - 4\gamma^2} + q(2n+1)$.

### 4.1 Real Potentials

An inspection to the discrete sequence of real spectra equation given by Eq.(27) shows that, first considering the real cases, i.e., all parameters $(V_0, q, \alpha)$ are real :

**( i )** Irrespective of the sign of $V_0$, for attractive or repulsive potential, bound states can exist. For any given $\alpha$ the spectrum consists of real eigenvalues $E_n(V_0, q, \alpha)$ depending on $q$ as shown in Fig.2.a,b. Fig.2.a shows that, while $V_0 \to 0$ in the ground state (i.e. $n = 0$), all energy curves are tending to the value $\approx 0.87\, mc^2$ for positive $q$ values and $1/\alpha = \lambda_c$, where $\lambda_c = \hbar/mc$ denotes the Compton wavelength of the KG particle. Otherwise, for the same value of $\alpha$ and negative $q$ values, when $V_0 \to 0$, all energy curves go to zero (Fig.2.b). Let us now discuss the limit of very short-ranged potential ($\alpha \to 0$). In this case the potential is close to the origin

$$V_q(x) \approx \frac{V_0}{q-1} + \frac{V_0}{(q-1)^2} \alpha\, x + O(\alpha^2 x^2) \qquad (28)$$

and behaves like a linear potential with a constant shift, $V_0/(q-1)$, where $\alpha$ denotes the range parameter and $V_0$, the coupling constant.

**( ii )** If $4\gamma^2 \leq q^2$, there exists bound states, otherwise there are no bound states.

**( iii )** If $V_0^2 \leq \dfrac{16 q^2 \gamma^2 (mc^2)^2}{4\gamma^2 + \kappa^2}$, there exists bound states, otherwise there are no bound states. Moreover, this relation which gives the critical coupling value leads to the result

$$n \leq \frac{\gamma}{qV_0}\sqrt{4q^2(mc^2)^2 - V_0^2} - \frac{1}{2q}\sqrt{q^2 - 4\gamma^2} - \frac{1}{2} \qquad (29)$$

i.e. there are only finitely many eigenvalues. In order that at least one level might exist, it is necessary that the inequality

$$q + \sqrt{q^2 - 4\gamma^2} \leq \frac{2\gamma}{V_0}\sqrt{4q^2(mc^2)^2 - V_0^2}$$

is fulfilled. As it can be seen from Fig.3.a,b, there are only two lower-lying states for the parameters $\alpha = mc/\hbar$, $q = \pm 1$.

**( iv )** If both conditions ( ii ) and ( iii ) are satisfied together, the bound states appear as seen from the energy expression (27). Fig.4 illustrates the ground state level as a function of the shape parameter $q$ for different values of the range parameter $\alpha$. While for $q < 0$ all of the negative bound states tend to zero asymptotically with decreasing $q$, for $q > 0$ and $1/\alpha > 0.5\lambda_c$ all of the positive bound states approach to the value $mc^2$.

For a more specific case $q = -1$, the generalized Hulthen potential is reduced to the shifted Woods-Saxon potential

$$V(x) = \frac{V_0}{1 + e^{-\alpha x}} - V_0\,, \qquad (30)$$

and then, its energy spectra yields

$$E_n = -\frac{V_0}{2} \pm \left[\sqrt{1 - 4\gamma^2} - (2n+1)\right]\sqrt{\frac{(mc^2)^2}{4\gamma^2 + \left[\sqrt{1-4\gamma^2} - (2n+1)\right]^2} - \frac{V_0^2}{16\gamma^2}}\,. \qquad (31)$$

Note that the expression (29) explains why there is a limitation on the number of bound states of Woods-Saxon potential (Fig.3.b). However, for any given $\alpha$, all of the eigenstates $E_n \leq 0$. Obviously, for given any $V_0$, as can be seen from Fig.5.a,b, all possible eigenstates have positive (negative) energies if $q$ parameter is positive (negative). It is almost note that there are some crossing points of the energy eigenvalues, i.e., there are the degenerations for the corresponding potential parameters.

( v ) For $q = 0$ although the potential reduces to the exponential potential

$$V(x) = -V_0 \, e^{-\alpha x}, \tag{32}$$

the eigenvalue expression (27) does not give an explicit form (when $q = 0$ is substituted in Eq.(27), it goes to infinity), i.e., the *NU*-method is not applicable to the exponential potential. Note that for this potential there is no an explicit form of the energy expression of bound states for Schrödinger [21], KG [19] and also Dirac equations [22].

Now let us discuss why the *NU*-method can not be applied to the exponential potential. In this case ( i.e. $q = 0$ ), avoiding to repeat the same development, we can write

$$\pi(z) = \pm \begin{cases} (i\gamma\, z + \mathcal{E}) & \text{for} \quad k = \beta^2 + 2i\gamma\, \mathcal{E} \\ (i\gamma\, z - \mathcal{E}) & \text{for} \quad k = \beta^2 - 2i\gamma\, \mathcal{E} \end{cases} \tag{33}$$

and the new function $\tau(z) = \tilde{\tau}(z) + 2\pi(z)$ yields

$$\tau(z) = (1 + 2\mathcal{E}) - 2i\gamma\, z \tag{34}$$

and then, we have another constant, $\lambda = k + \pi'$, given by

$$\lambda = \beta^2 - 2i\gamma\, \mathcal{E} - i\gamma. \tag{35}$$

It seems from the last equation that if and only if $i\gamma$ is real, $\lambda$ is a real constant. Hence, in order to apply the *NU*-method to this potential $\gamma = V_0/\hbar c \alpha$ should be imaginary! This leads us to the result that either $V_0$ or $\alpha$ must be imaginary. So, we think that this is an open problem nowadays.

**4.2 Bound States of s-wave pions**

If the depth of the potential is set to $V_0 = \dfrac{Ze^2}{R}$, where the nuclear radius is $R = r_0 A^{1/3}$ with $r_0 \approx 1.2 \, fm$ and the number of nucleons $A \approx 2.5\, Z$; the critical coupling value leads to the results

$$Z \leq 22 \quad \text{for} \quad q = 0.10$$
$$Z \leq 88 \quad \text{for} \quad q = 0.25$$
$$Z \leq 250 \quad \text{for} \quad q = 0.50$$

i.e., there are no bound states solutions for the *s*-wave pions ($\pi^- - meson$) with the Compton wavelength [24] $\lambda_\pi = \hbar/m_\pi c$ in the ground state, because the energy expression which is given by Eq.(27) becomes imaginary. When $q = 0.10$, the energy eigenvalue for the *s*-wave pions at $Z = Z_{critical} = 22$ are $E_0 \cong 0.946\, m_\pi c^2$ (Fig.6). The corresponding binding energy is about $E_b = E_0 - m_\pi c^2 \cong -7.54\, MeV$. At $Z = 16$, $E_0$ reaches the maximum value $m_\pi c^2$.

### 4.3 Complex potentials

Now let us consider cases ; namely at least one of the parameters is imaginary or complex.

( i ) Let $V_0$ and $q$ be complex parameters, i.e., $V_0 = V_{0R} + iV_{0I}$ and $q = q_R + iq_I$, where $V_{0R}$, $V_{0I}$, $q_R$ and $q_I$ are arbitrary real parameters and $i = \sqrt{-1}$. So, one can find different sets of parameters for this condition. In this case, if $\text{Re}(V_0) = 0$ and $\text{Re}(q) = 0$, then the potential transforms to the form

$$V_q(x) = V_{0I} \frac{q_I\left[2\cosh^2(\alpha x) - \sinh(2\alpha x) - 1\right] - i\left[\cosh(\alpha x) - \sinh(\alpha x)\right]}{1 + q_I^2\left[2\cosh^2(\alpha x) - \sinh(2\alpha x) - 1\right]} \qquad (36)$$

and hence, if and only if $V_{0I}^2(4\gamma^2 + \kappa^2) \leq 16 q_I^2 \gamma^2 (mc^2)^2$, it has a real spectra.

( ii ) If $\alpha$ is a complex parameter, $\alpha = \alpha_R + i\alpha_I$, it also has a real spectra. In this case, for $\alpha_R = 0$, i.e., $\alpha$ is a completely imaginary parameter, such potentials are written as a complex function

$$V_q(x) = V_0 \frac{q - \cos(\alpha_I x) + i\sin(\alpha_I x)}{q^2 - 2q\cos(\alpha_I x) + 1} \qquad (37)$$

which is *PT*-symmetric but non-Hermitian. It has a real spectra if and only if $16 q^2 \gamma^2 (mc^2)^2 \leq V_0^2 \left\{ 4\gamma^2 - \left[\sqrt{q^2 + 4\gamma^2} + q(2n+1)\right]^2 \right\}$.

( iii ) It is interesting to note that all three parameters $V_0$, $q$ and $\alpha$ are altogether imaginary, in this case we obtain the potential as

$$V_q(x) = V_{0I} \frac{q_I - \sin(\alpha_I x) - i\cos(\alpha_I x)}{q_I^2 - 2q_I \sin(\alpha_I x) + 1}. \qquad (38)$$

This form of the potential is non-*PT*-symmetric and also non-Hermitian but it has a real spectra if and only if $16 q_I^2 \gamma^2 (mc^2)^2 \leq V_{0I}^2 \left\{ 4\gamma^2 - \left[\sqrt{q_I^2 + 4\gamma^2} + q_I(2n+1)\right]^2 \right\}$.

## 5. Eigenfunctions

Let us now find the corresponding wave functions. In the *NU*-method the wave function is constructed as a multiple of two independent parts

$$\psi(z) = \phi(z) y(z) \tag{39}$$

where $y(z)$ is the polynomial solution of hypergeometric-type equation which is described with a weight function [15]. The other part is defined as a logarithmic derivative function as

$$\phi'(z)/\phi(z) = \pi(z)/\sigma(z). \tag{40}$$

By substituting $\pi(z)$ and $\sigma(z)$ in Eq.(40) and then solving the first order differential equation, one can find

$$\phi(z) = z^{\varepsilon} (1 - qz)^{(a+q)/2q}. \tag{41}$$

It is easy to find the other part of the wave function from the definition of weight function as

$$\rho(z) = z^{2\varepsilon} (1 - qz)^{a/q}, \tag{42}$$

and substituting in the Rodrigues relation, we get

$$y_{nq}(z) = B_{nq} z^{-2\varepsilon} (1 - qz)^{-a/q} \frac{d^n}{dz^n} \left[ z^{n+2\varepsilon} (1 - qz)^{n+(a/q)} \right]. \tag{43}$$

Combining all of the results, the unnormalized *s*-wave functions can be constructed as

$$\psi_{nq}(z) = B_{nq} z^{-\varepsilon} (1 - qz)^{(q-a)/2q} \frac{d^n}{dz^n} \left[ z^{n+2\varepsilon} (1 - qz)^{n+(a/q)} \right]. \tag{44}$$

In the limit $q \to 1$, the eigenfunctions are obtained in terms of Jacobi polynomials, which is one of the orthogonal polynomials, giving $y_{nq}(z) \approx P_n^{(2\varepsilon, b)}(1 - 2z)$ whereas the unnormalized *s*-wave functions are

$$\psi_n(z) = C_n z^{\varepsilon} (1 - z)^{(1+b)/2} P_n^{(2\varepsilon, b)}(1 - 2z) \tag{45}$$

with $z = e^{-\alpha x}$, $b = \sqrt{1 - 4\gamma^2}$. Obviously, in this case, the eigenvalues revert that of the standard Hulthen potential.

## 6. Results and Discussion

We have solved the Klein-Gordon equation for the generalized Hulthen potential in relativistic quantum mechanics. According to the second condition of complex quantum mechanics [1], the eigenfunctions of the cases obtained by $\alpha \to i\alpha$ are simultaneously eigenstates of *PT* operator. As it should be expected (see Eq.(27)), for any given set of potential parameters $\alpha$ and $V_0$, although the energy levels of Woods-Saxon potential ($q = -1$) are negative, the energy levels of standard Hulthen potential ($q = 1$) are positive. On the other hand, we can state that our results are not only be interesting for pure theoretical physicist but also for experimental physicist, because the results are exact and more general.

As an example, an application to the pionic systems is given. We have already mentioned that we have found some simple relations among the potential parameters for bound states. We show that, it is possible to get relativistic bound states of complex quantum mechanical formulation. Thus the relativistic bound state spectra of the generalized Hulthen potential exhibit the effects of a well of depth $V_0$, potential range parameter $\alpha$, and the shape parameter $q$.

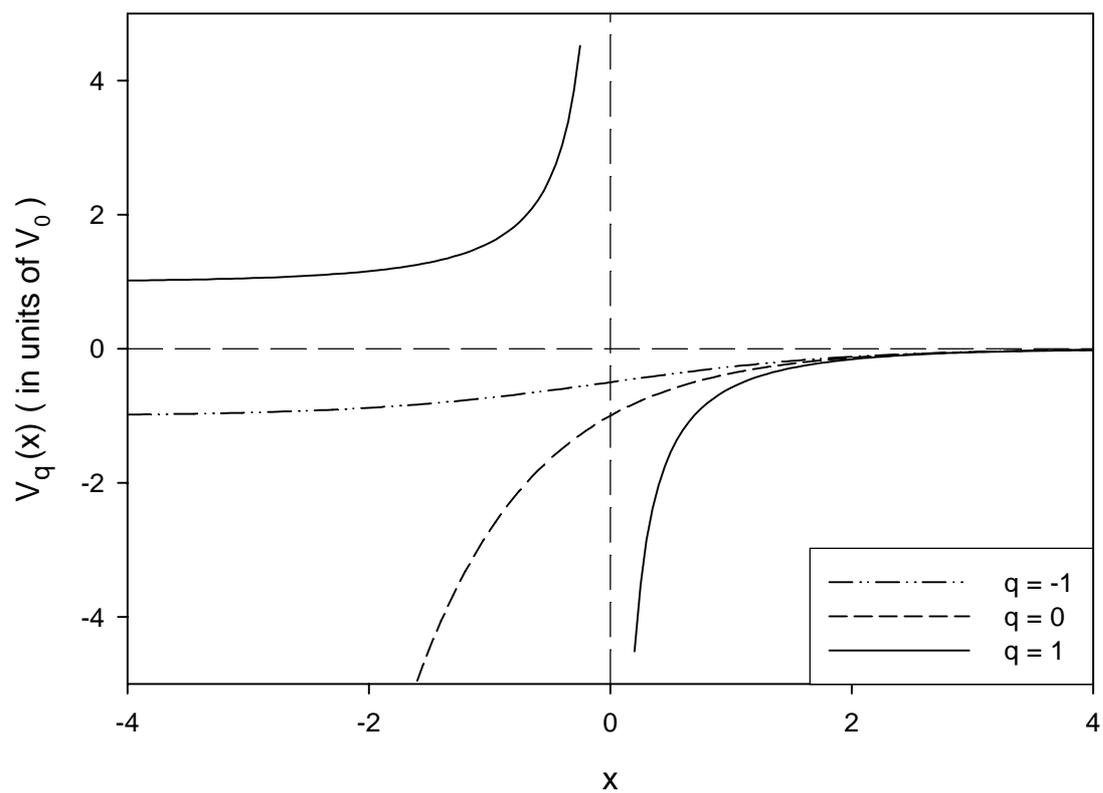

**Fig.1.a.** A schematical representation of the generalized Hulthen potential for three different values of the shape parameter $q$ for $\alpha = 1$.

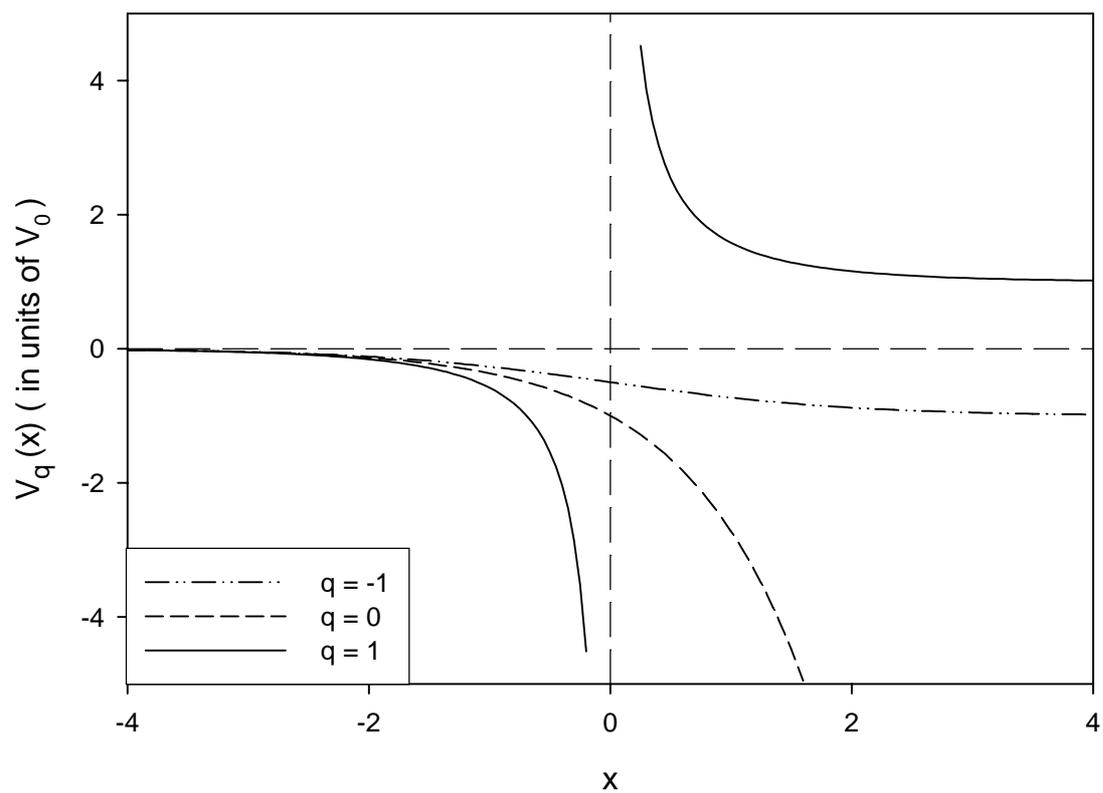

**Fig.1.b.** A schematical representation of the generalized Hulthen potential for three different values of the shape parameter $q$ for $\alpha = -1$.

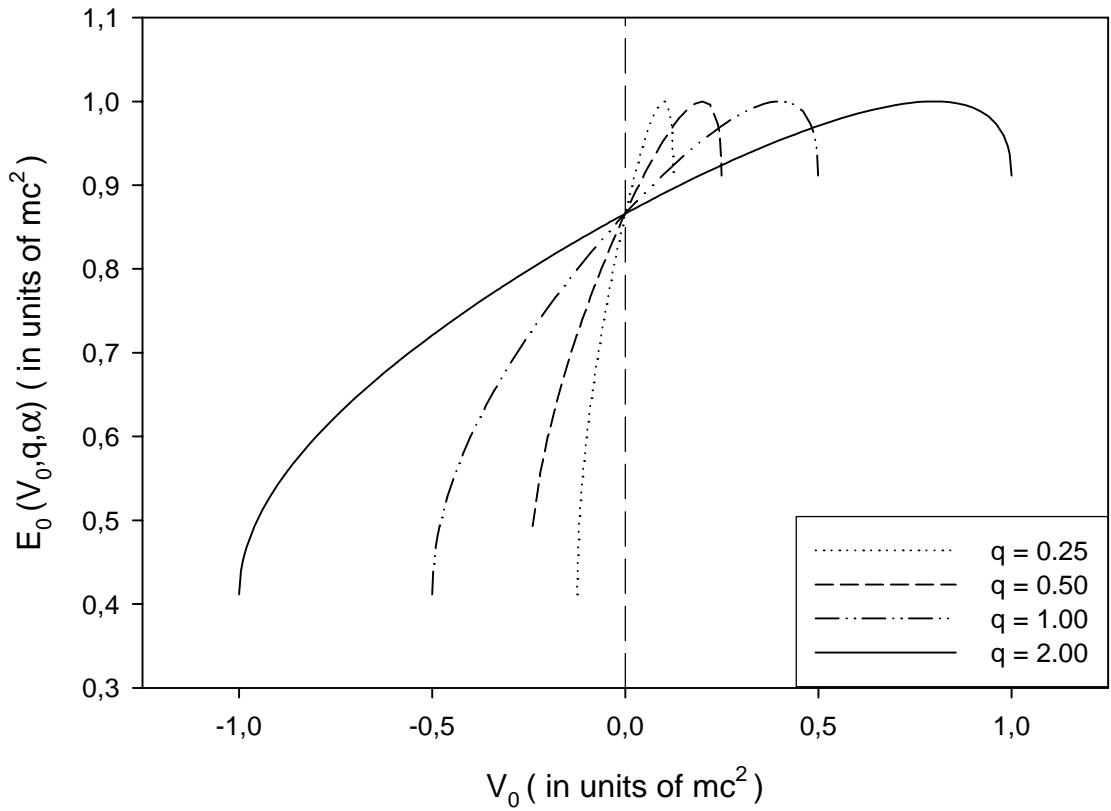

**Fig.2.a.** The variation of the ground state energy of a Klein-Gordon particle with respect to the coupling constant $V_0$ for the positive shape parameters of the generalized Hulthen potential where $\alpha = mc/\hbar$.

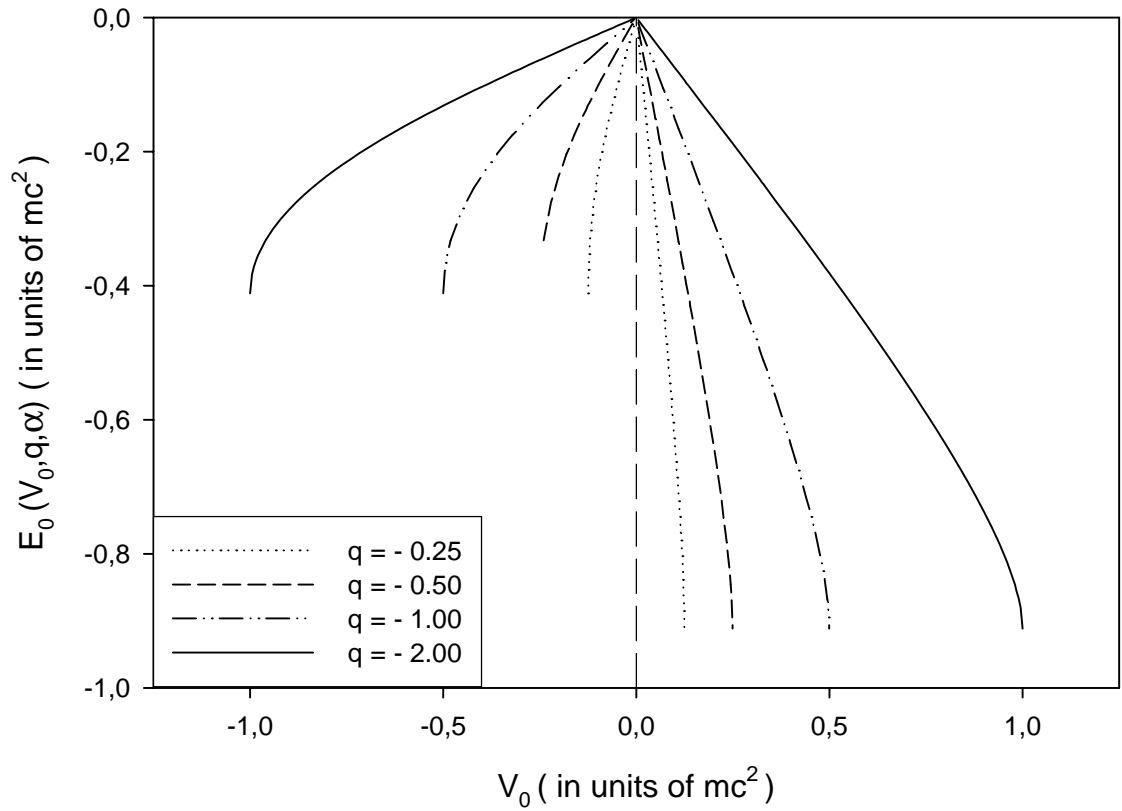

**Fig.2.b.** The variation of the ground state energy of a Klein-Gordon particle with respect to the coupling constant $V_0$ for the nagative shape parameters of the generalized Hulthen potential where $\alpha = mc/\hbar$.

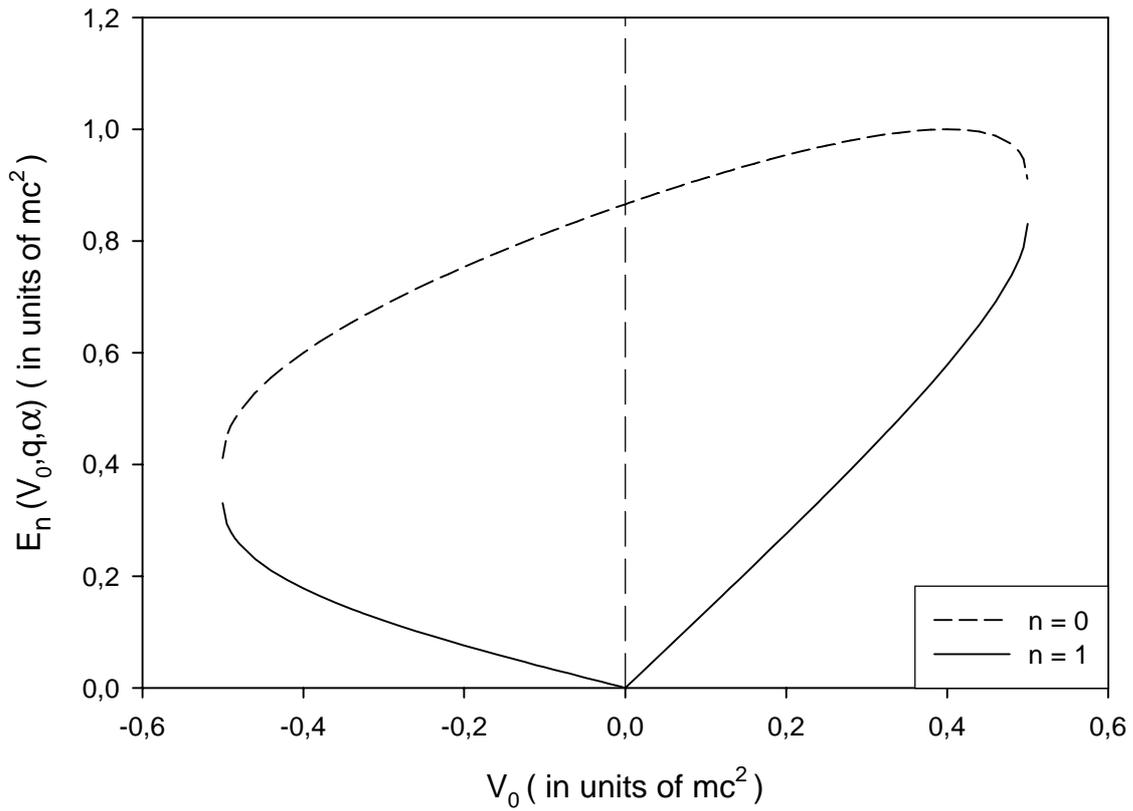

**Fig.3.a.** The variation of the energy eigenvalues with respect to the coupling constant $V_0$ for the standard Hulthen potential ($q=1$) where $\alpha = mc/\hbar$.

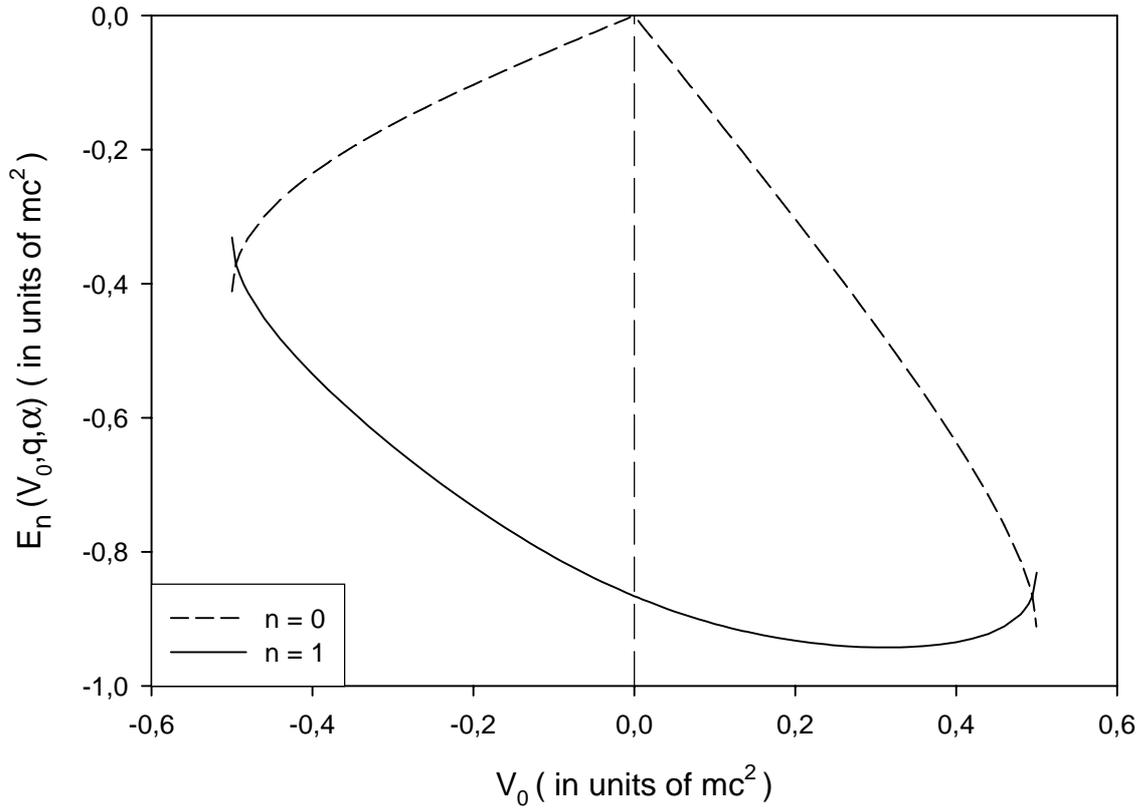

**Fig.3.b.** The variation of the energy eigenvalues with respect to the coupling constant $V_0$ for the shifted Woods-Saxon potential ($q = -1$) where $\alpha = mc/\hbar$ has been choosen. There are the degenerations at the values $V_0 = \pm 0.495\, mc^2$ of the coupling constant.

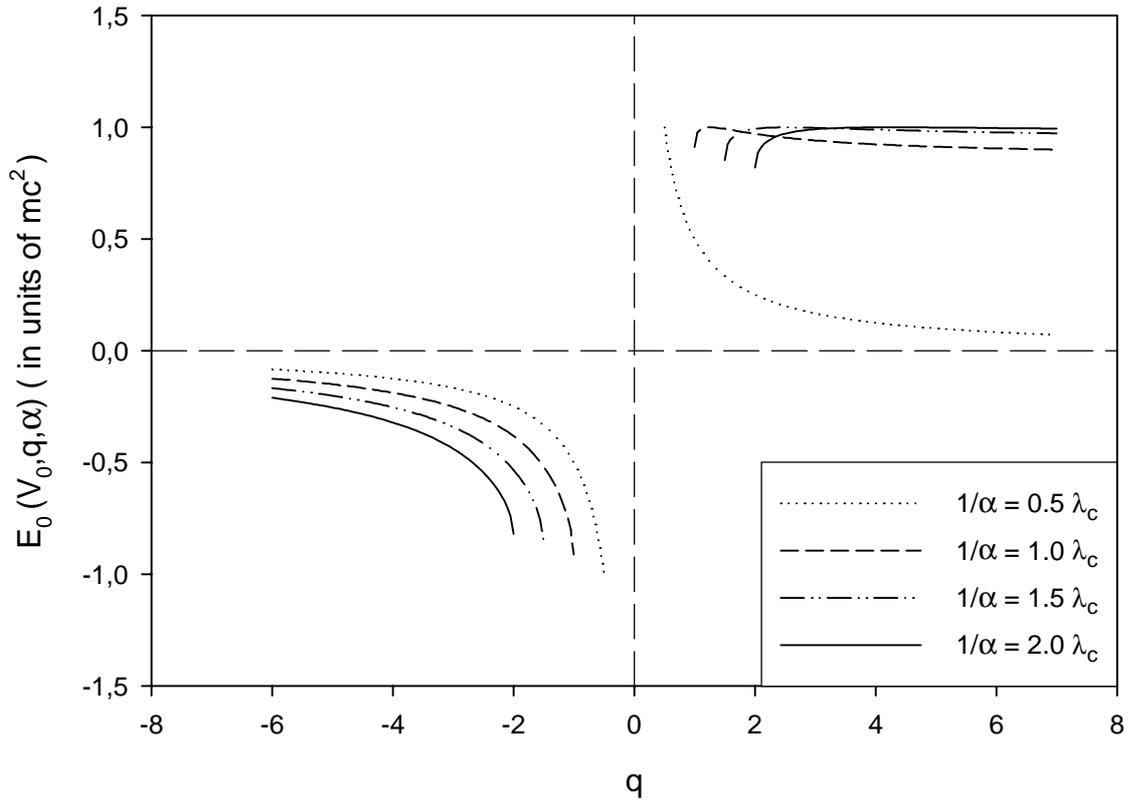

**Fig.4.** The ground state eigenvalues as a function of the shape parameter $q$ for different values of the range parameter $\alpha$ where $V_0 = 0.5\,mc^2$.

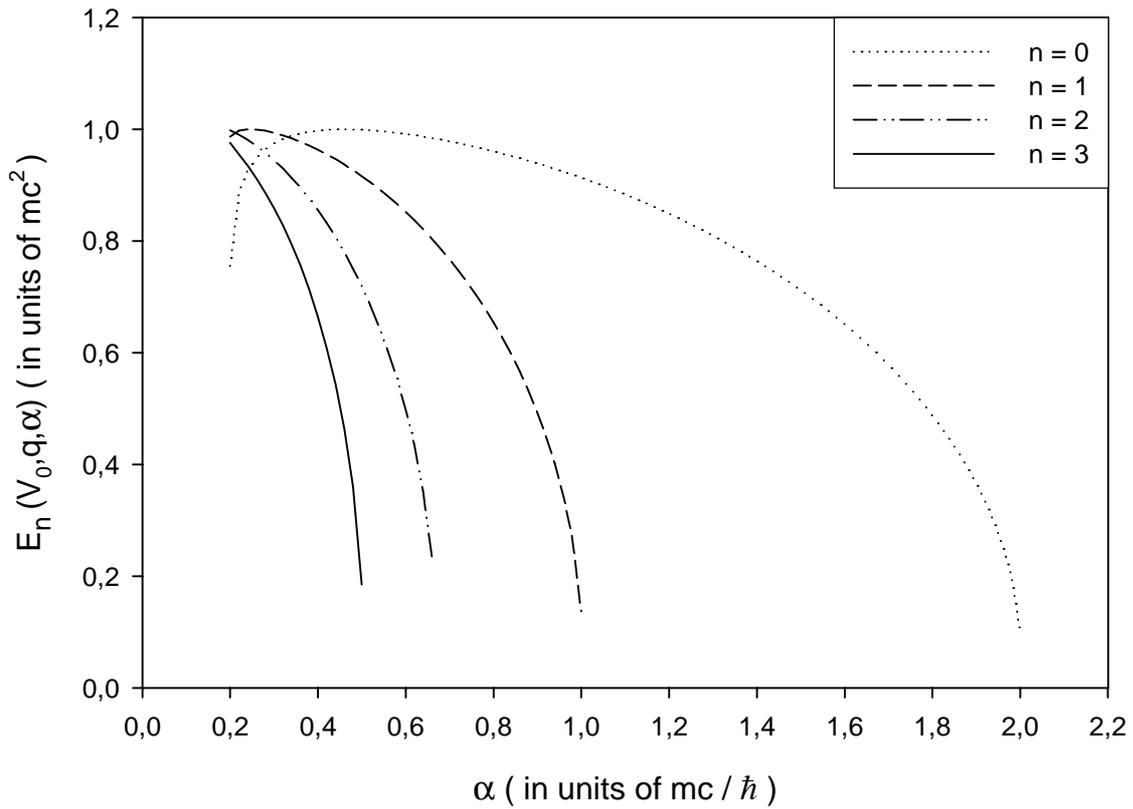

**Fig.5.a.** The variation of the energy eigenvalues with respect to the range parameter $\alpha$ for the standard Hulthen potential ($q=1$) where $V_0 = 0.1\,mc^2$ has been choosen. The curves are plotted for the first four values of the quantum number *n*.

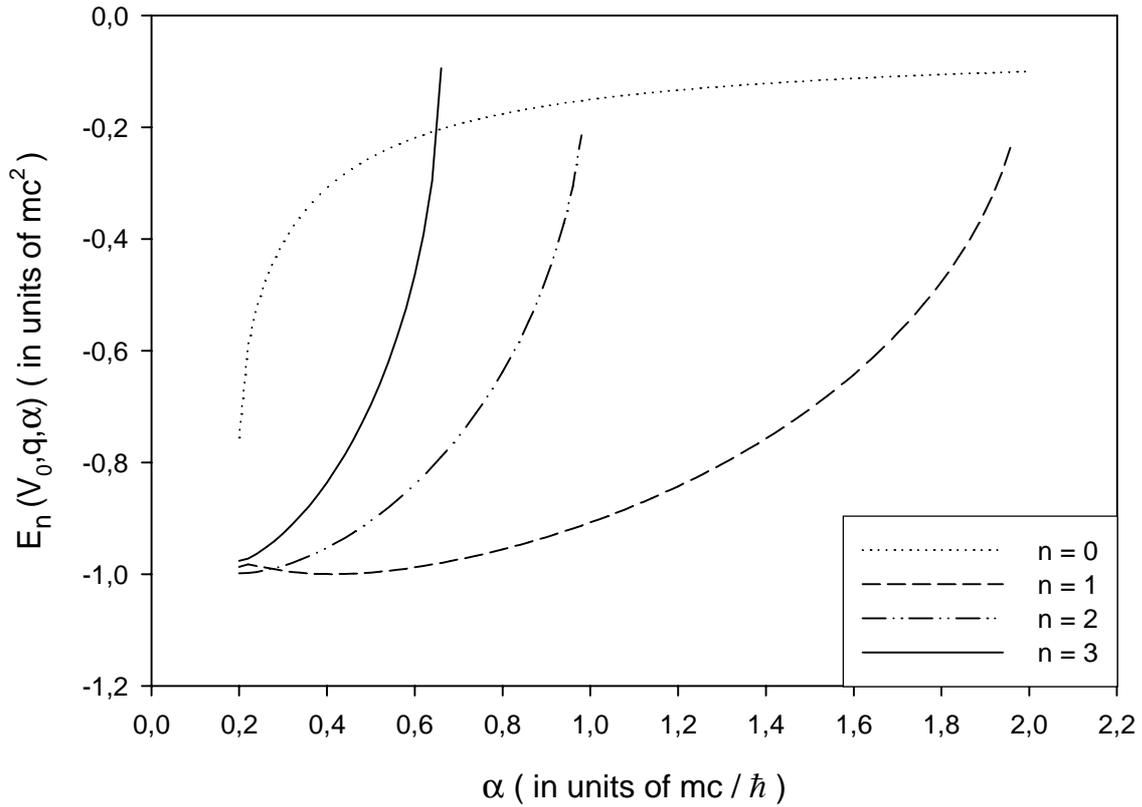

**Fig.5.b.** The variation of the energy eigenvalues given by Eq.(31) of a Klein-Gordon particle with respect to the range parameter $\alpha$ for the shifted Woods-Saxon potential ($q = -1$) where $V_0 = 0.1\, mc^2$ has been choosen. The curves are plotted for the first four values of the quantum number *n*.

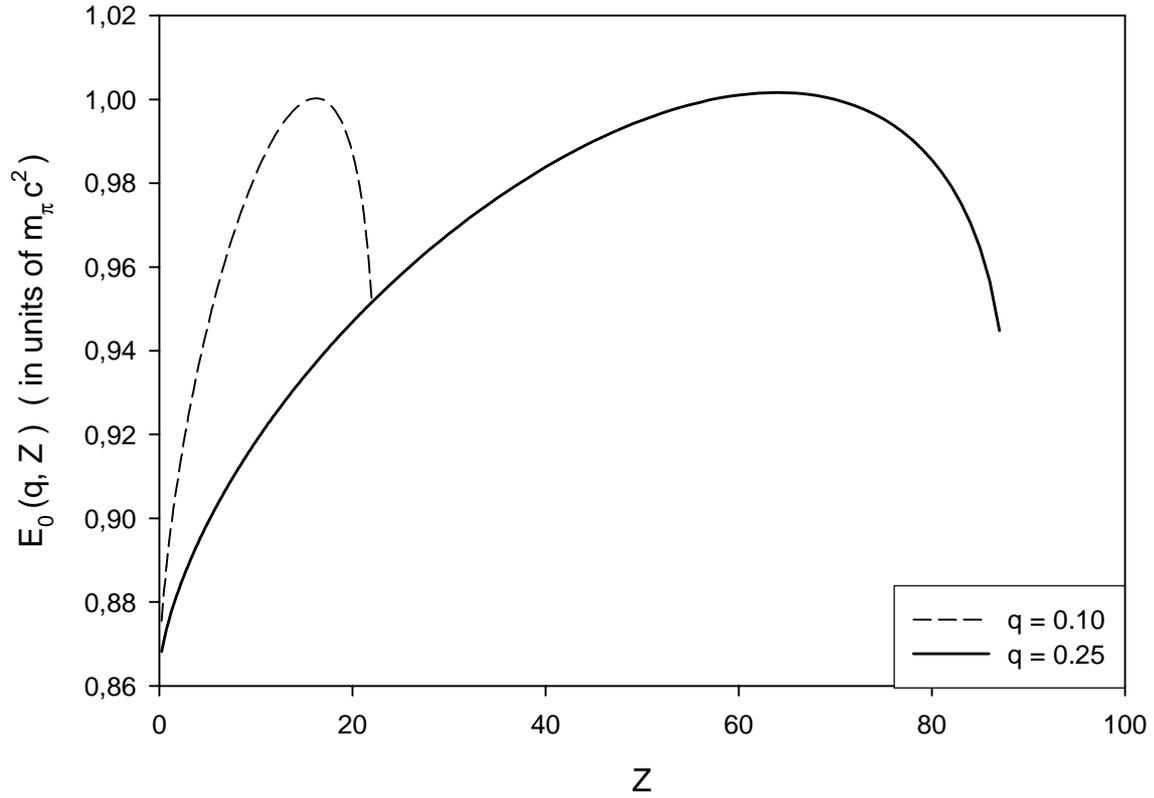

**Fig.6.** The ground state eigenvalues of a $\pi^-$ – meson in the generalized Hulthen potential as a function of the nuclear charge $Z$. The constant $\alpha$ characterizing the range of the potential is $1/\alpha = \lambda_\pi = \hbar/m_\pi c$. When $q = 0.25$, there are no solutions for $Z > 88$.